\documentclass{evn2002}
\newcommand{\bm}[1]{{\mbox{\boldmath$#1$}}}
\begin{document}
\title{General relativistic model for experimental measurement of the speed
of propagation of gravity by VLBI}

\author{S. Kopeikin\inst{1,3} and E. Fomalont\inst{2}}

\institute{Department of Physics and Astronomy, University of Missouri -
Columbia, 223 Physics Bldg., Columbia, Missouri, 65211, USA
\and National Radio Astronomical Observatory, 520 Edgemont Road,
Charlottesville, VA 22903, USA
 \and E-mail: kopeikins@missouri.edu}

\abstract
{A relativistic sub-picosecond model of gravitational time delay in radio
astronomical observations is worked out and a new experimental test of
general relativity is discussed in which the effect of retardation of gravity
associated with its finite speed can be observed.  As a consequence, the
speed of gravity can be measured by differential VLBI observations.
Retardation in propagation of gravity is a central part of the Einstein
theory of general relativity which has not been tested directly so far. The
idea of the proposed gravitational experiment is based on the fact that
gravity in general relativity propagates with finite speed so that the
deflection of light caused by the body must be sensitive to the ratio of the
body's velocity to the speed of gravity. The interferometric experiment can
be performed, for example, during the very close angular passage of a quasar
by Jupiter. Due to the finite speed of gravity and orbital motion of Jupiter,
the variation in its gravitational field reaches observer on Earth not
instantaneously but at the retarded instant of time and should appear as a
velocity-dependent excess time delay in addition to the well-known Shapiro
delay, caused by the static part of the Jupiter's gravitational field. Such
Jupiter-QSO encounter events take place once in a decade. The next such event
will occur on September 8, 2002 when Jupiter will pass by quasar J0842+1835
at the angular distance $3.7'$. If radio interferometric measurement of the
quasar coordinates in the sky are done with the precision of a few
picoseconds ($\sim$ 5 $\mu$as) the effect of retardation of gravity and its
speed of propagation may be measured with an accuracy about 10\%.}
\date{}
\titlerunning{Relativistic model for experimental measurement of the speed of
gravity }
\maketitle

%

\section{Theoretical Background}
Experimental verifications of the basic principles underlying
Einstein's general relativity theory are important for fundamental
physics. All previous experimental tests of general relativity in
the solar system have relied upon the static Schwarzchild solution
(\cite{will}) and, therefore, were not sensitive to the effects
entirely associated with the propagation speed of gravity.
It is worth noting that gravitational waves are inherent to the radiative
(far) zone of a system emitting the waves (\cite{mtw}; \cite{ligo}). However, the
gravitational waves do not propagate freely through the interior of a
non-radiative (near) zone of the system. Nevertheless, the process of
generation of gravitational waves produces retarded effects in the near zone
leading to appearance of the gravitational radiation reaction force in the
relativistic equations of motion of extended bodies comprising a
self-gravitating astronomical system (\cite{dgks89}). Existence of this force
is a consequence of the finite speed of propagation of gravity as it was
experimentally confirmed by Taylor (1994).

We have found (\cite{kopajl}) that the gravitational bending of
light  passing through the gravitational field of a moving massive
object like Jupiter, though being dominated by the
spherically-symmetric component of its gravitational field, also
contains terms associated with the finite speed of propagation of
gravity.
Our calculations
reveal that electromagnetic signals interact with the light-ray
deflecting bodies only through the retarded gravitational fields
-- the observational effect which must be accounted for in precise data
processing algorithms adopted for the microarcsecond space astrometry.

This gravitational light-propagation theory, in the case of a static
spherically-symmetric field, gives the same result as that
predicted by Einstein for the bending of light, $\alpha_E\simeq
4GM/(c^2R\theta)$, where $M$ is the mass of the light-ray deflecting
body, $R$ is the distance from observer to the body, and $\theta$
is the (small) angle in the sky between the undisturbed geometric
positions of the source of light and the center of mass of the
massive body.  Furthermore, this theory allows to calculate the
correction, $\alpha_{PG}$, to the Einstein deflection $\alpha_E$ related to
the variability of the gravitational field produced by the motion
of the light-ray deflecting body. We were successful in proving
(\cite{kopajl}) that these corrections in the bending of light are
inherently associated with the finite speed of propagation of
gravity and in case of slowly moving bodies can be parameterized as
$\alpha_{PG}\simeq
(1+\delta)(\alpha_E/\theta)(v/c)$, where $v$ is the orbital
velocity of the light-ray deflecting body with respect to the
barycenter of the solar system projected on the plane of the sky,
and $\delta=c_g/c-1$ is a fitting parameter used in data analysis. It is
chosen such that
$\delta=0$, if the speed of gravity $c_g$ equals the speed of
light $c$.

Parameter $\delta$ is a close analogue of the parameter
$\alpha_2=(c_g/c)^2-1$ of the parameterized post-newtonian (PPN) formalism
which quantifies possible violation of the local Lorentz invariance
(\cite{will})\footnote{One notices that $\delta=\alpha_2/2$ in the first
approximation.}. It was shown (\cite{nord}) that $\alpha_2<4\times 10^{-7}$
under the (rather restrictive) assumption that the preferred frame is
realized by the cosmological Hubble flow. If one abandons any anthropic
assumption about the speed of the solar system with respect to an (actually
unknown) preferred frame the limit on $\alpha_2< 0.1$ can be obtained from
the analysis of the anomalous perihelion shifts of inner planets
(\cite{will}). The primary purpose of our experiment is, however, to observe
directly the effect of retardation in
propagation of gravitational field in the solar system
rather than improving limits on $\alpha_2$. Nevertheless, we would like to
emphasize that relativistic effects in propagation of light through time
dependent gravitational fileds are also sensitive to violation of local
Lorentz invariance.

The largest measurable contribution to the variable,
time-dependent part of the solar system gravitational field comes out
from the orbital motion of Jupiter. The minimal value of
the impact parameter of an incoming light ray from a quasar that can be
achieved
for Jupiter is also the least possible amongst all the solar system bodies.
Therefore, it is highly sensible to make an attempt for
detection of the ``gravity retardation" effect by observing very accurately
the gravitational
deflection of light from a background source (quasar) caused by
the motion of Jupiter around the barycenter of the solar system.
As explained in (\cite{kopajl}) the magnitude of the observed effect
is directly translated to the measured value of the propagation speed of
gravity $c_g$. This is the essence of the new test of general
relativity which has never been done before with sufficient accuracy.

Radio astronomical methods of VLBI are the most accurate for
measuring gravitational deflection of electromagnetic waves. The
two most precise measurements of the bending of radio waves near
the sun (\cite{leb}; \cite{rob}) were accurate to about 0.1\%.
However, these observations were insensitive to the speed of gravity effect
$\alpha_{PG}$
because of the relatively large impact parameter of the incoming
light ray. Even at the solar limb, the magnitude of $\alpha_{PG}$ is $\sim
10^{-5}$ of the
static gravitational bending $\alpha_E=1.75'$ and is totally
unobservable because of the highly turbulent solar magnetosphere.

On September 8, 2002 Jupiter will pass at an angular distance of
$3.7'$ from the quasar J0842+1835 making an ideal
celestial configuration for measuring the speed of propagation of
gravity by using the phase-referencing VLBI technique (\cite{fk2002}). The
encounter in 2002 is especially favorable because: (1) it
occurs when Jupiter is relatively far from the Sun (the next
near occultation which occurs is a few degrees from the Sun), and (2)
the five critical hours of the closest approach occur when Jupiter is near
the transit line
for VLBA observations.

 We have
estimated that for this Jupiter-quasar encounter the
deflection from the static gravitational field and from the
propagation of gravity are, respectively, $\alpha_E=1.26$ mas and
$\alpha_{PG} =
53$ $\mu$as \footnote{It corresponds to the time delays 122.2 and 5.1
picoseconds respectively on a baseline $b=6000$ km.} both in the plane of the
sky with the static bending
radially from Jupiter and the propagation bending in the
direction of Jupiter's motion (see Eqs. (\ref{a6}) and (\ref{a7}) in Sec. 3).
As one can see the ratio
$|\alpha_{PG}|/\alpha_E\simeq 0.04$ is much larger for Jupiter than for
the Sun which is explained by the ability to get a smaller impact
parameter $\theta$ for the light ray passing by Jupiter than that
for the Sun.

\section{Relativistic Model of VLBI Time Delay}

Detection of the effect of gravity propagation requires a more advanced VLBI
model for light propagating in the time dependent gravitational field of the
solar system. Such a model must be valid to a precision of better than 0.1
ps. In the present paper we discuss the appropriate corrections to the
standard Shapiro time delay (\cite{shap}) which bring the accuracy of the
model up to the necessary threshold.

The general formula for the relativistic time delay $\Delta T$ in the field of
a system of moving bodies is given in (\cite{kopajl})
\begin{eqnarray}
\label{3}
\Delta T&=&(1+\gamma){G\over c^3}\sum_{a=1}^N
m_a\;\displaystyle{\int^{s}_{s_{0}}}\frac{\left(1-{1\over c}{\it\bm
k}{\bm\cdot}{\it\bm v}_a(\zeta)\right)^2\Upsilon(\zeta)\,d\zeta}{t^{\ast}-
\zeta+{1\over c}{\it\bm
k}{\bm\cdot}{\it\bm x}_a(\zeta)}
,
\end{eqnarray}
where $\Upsilon(\zeta)=1/\sqrt{1-c^{-2}v^2_a(\zeta)}$ is the Lorentz factor,
$\gamma$ is the PPN parameter (\cite{will}), $m_a$ is the mass of the $a$th
body, $t^*$ is the time of the closest approach of electromagnetic
signal to the barycenter of the Solar system \footnote{The time
$t^*$ is used in calculations as a mathematical tool only. It has no real
physical meaning because of its dependence on the choice of a coordinate
system.}, ${\it\bm x}_a(t)$ are coordinates of the
$a$th body, ${\it\bm v}_a(t)=d{\it\bm x}_a(t)/dt$ is the
(non-constant) velocity of the $a$th light-ray deflecting body,
${\it\bm k}$ is the unit vector from the point of emission to the
point of observation, $s$ is a retarded time obtained by solving
the gravitational null cone equation for the time of observation
of photon $t=s+c_g^{-1}|{\it\bm x}-{\it\bm x}_a(s)|$, and $s_0$ is found by
solving the same equation
written down for the time of emission of the photon $t_0=s_0+c_g^{-1}|{\it\bm
x}_0-{\it\bm x}_a(s_0)|$.

One emphasizes that the equations for the retarded times depend on the speed
of gravity $c_g$, but not the speed of light $c$. This is because they were
obtained by solving Einstein equations for the space-time metric
perturbations by making use of retarded Lienard-Wiechert tensor potentials
(\cite{ks}). These retarded gravitational potentials describe propagation of
gravity without any relation to the problem of propagation of light in the
gravitational field. In general relativity $c_g=c$ numerically. However, when
light propagates through time-dependent gravitation field, in principle, one
can separate relativistic effects associated with propagation of light and
gravity.

The Earth
and Sun also contribute significantly to the gravitational time
delay and must be included in the data processing algorithm in
order to extract accurately the effect of retardation of gravity.
Precise calculation of the integral (\ref{3}) for two radio antennas gives the differential VLBI time delay 
\begin{eqnarray}
\label{29}
\Delta&=&\Delta_2 T-\Delta_1 T=\Delta_\oplus+\Delta_\odot+\Delta_J+\Delta_{JPG}\;.
\end{eqnarray}
The first term in the right hand side of (\ref{29}) describes the
gravitational (Shapiro) time delay due to the gravitational field
of the Earth
\begin{eqnarray}
\label{a1}
\Delta_\oplus&=&(1+\gamma){GM_\oplus\over c^3}
\ln\frac{X_{1}+{\bm K}\cdot{\bm X}_{1}} {X_{2}+{\bm K}\cdot{\bm
X}_{2}}\;.
\end{eqnarray}
It can reach 21 ps for the baseline $b=6000$ km. 

The second term in the right hand side of (\ref{29}) describes the gravitational (Shapiro) time
delay due to the Sun
\begin{eqnarray}
\label{a2}
\Delta_\odot&=&(1+\gamma){GM_\odot\over c^3}
\ln\frac{r_{1\odot}+{\bm K}\cdot{\bm r}_{1\odot}} {r_{2\odot}+{\bm
K}\cdot{\bm r}_{2\odot}}\;.
\end{eqnarray}
It can vary (for $b=6000$ km) from $17\times 10^4$ ps for the light ray
grazing the Sun's limb to only 17 ps when direction to the source of light is
opposite to the Sun. 

The third term in
the right hand side of (\ref{29}) is the Shapiro time delay due to
the static part of the gravitational field of Jupiter
\begin{eqnarray}
\label{a3}
\Delta_J&=&(1+\gamma){GM_J\over c^3}\left(1+{\bm K}\cdot{\bm
v}_J\right) \ln\frac{r_{1J}+{\bm K}\cdot{\bm r}_{1J}} {r_{2J}+{\bm
K}\cdot{\bm r}_{2J}}\;.
\end{eqnarray}

Finally, the forth term in the
right hand side of (\ref{29}) is the time delay caused by the finite speed of
gravity as predicted by general relativity theory (\cite{kopajl})
\begin{eqnarray}
\label{a4}
\Delta_{JPG}&=&2(1+\delta){GM_J\over c^4} {{\bm b}\cdot{\bm
v}_J+({\bm b}\cdot{\bm N}_{1J})({\bm K}\cdot{\bm v}_{J})\over
r_{1J}+{\bm K}\cdot{\bm r}_{1J}}\;.
\end{eqnarray}
In formulas (\ref{a1})--(\ref{a4}) we use the following notations: $M_\oplus$
-- mass of the Earth, $M_\odot$ -- mass of the
Sun, $M_J$ -- mass of Jupiter, ${\bm v}_J(t_1)$ -- the barycentric velocity
of Jupiter, and ${\bm K}$ -- the unit vector from the barycenter of the solar
system to the quasar observed. Also, for each $i=1,2$, one has the baseline
vector ${\bm b}={\bm X}_1-{\bm X}_2$, $r_{i\odot}=|{\bm r}_{i\odot}|$,
$r_{iJ}=|{\bm r}_{iJ}|$, ${\bm N}_{1J}={\bm r}_{1J}/r_{1J}$, ${\bm
r}_{i\odot}={\bm x}_i(t_i)-{\bm x}_\odot(t_i)$, ${\bm r}_{iJ}={\bm
x}_i(t_i)-{\bm x}_J(t_i)$, ${\bm x}_i={\bm x}_\oplus(t_i)+{\bm X}_i(t_i)$,
where ${\bm X}_i(t_i)$ are the geocentric coordinates of $i$-th
VLBI station, ${\bm x}_\oplus$ -- the barycentric coordinates of the
geocenter,
${\bm x}_\odot$ --
barycentric coordinates of the Sun, ${\bm x}_J$ -- barycentric
coordinates of Jupiter, and $t_i$ is time of arrival of the plane front of
electromagnetic wave from quasar to the $i$th VLBI station.

We notice there are two relativistic parameters to be measured in
order to test validity of general relativity theory - the PPN
parameter $\gamma$, and the speed of gravity parameter
$\delta=c_g/c-1$. The best experimental measurement of parameter
$\gamma$ had been conducted by Lebach et al. (1995) who
obtained $\gamma=0.9996\pm 0.0017$ in an excellent agreement with
general relativity. The primary goal of the new experimental test of general
relativity is to measure the
parameter $\delta$ which will set up limits on the numerical
value of the speed of gravity $c_g$ (\cite{fk2002}).

During the passage of Jupiter near the quasar the time-dependent
impact parameter ${\bm{\xi}}(t)$ of the light ray with respect to
Jupiter will be always small as compared with the distance from
Earth to Jupiter which will be approximately 6 AU. It is convenient to
introduce the unit vector ${\bf n}={\bm{\xi}}/|{\bm{\xi}}|$
along the direction of the impact parameter according to
definition $
\sin\theta\;{\bf{n}}=({\bf K}\times({\bf
N}_{1J}\times{\bf K})),
$
where $\theta$ is a small
angle between the unperturbed astrometric position of the quasar
and that of Jupiter. Making use of the (impact parameter $\theta$)
expansion ${\bf N}_{1J}=\;-(1-\theta^2/2){\bf
K}+\theta\,{\bf{n}}+O(\theta^3),
$
we obtain the functional structure of the Shapiro time delay $\Delta_J$ and
the speed of gravity delay $\Delta_{JPG}$ in a more explicit form
(assuming for simplicity $\gamma=1$)
\begin{eqnarray}
\label{a6}
\Delta_J&=&{4GM_J\over
c^3r_{1J}}\left[{{\bm n}\cdot{\bm B}\over \theta}+{({\bm n}\cdot{\bm
B})^2\over r_{1J}\theta^2}-{({\bm K}\times{\bm B})^2\over
2r_{1J}\theta^2}\right]\;,
\\\nonumber\\\label{a7}
\Delta_{JPG}&=&(1+\delta){4GM_J\over
c^4r_{1J}}{{\bm b}\cdot{\bm v}_J-({\bm K}\cdot{\bm v}_J)({\bm
K}\cdot{\bm b})\over \theta^2}\;,
\end{eqnarray}
where $
{\bm B}={\bm b}-c^{-1}({\bm K}\cdot{\bm b})({\bm v}_2-{\bm v}_J)+O(c^{-2})\;,
$
and all quantities in the right sides of Eqs. (\ref{a6})--(\ref{a7}) are
taken at the time $t_1$.
\section{The Effect of the Magnetosphere of Jupiter}

In addition to various special and general relativistic effects in the time
of propagation of electromagnetic waves from the quasar to the VLBI antenna
network, we must account for the effects produced by the jovian
magnetosphere. Measurements obtained during the occultations of Galileo by
Jupiter indicate (\cite{jupmag}) that near the surface of Jupiter the
electron plasma density reaches the peak
intensity $N_0 = 1.0\times 10^{10}$ m$^{-3}$. We shall assume that the jovian
magnetosphere is spherical \footnote{In reality the magnetosphere has a dipole
structure and we speculate that our
model which assumes circular symmetry grossly underestimates the
plasma content along the polar direction where the closest approach
occurs.} and a radial drop-off of the plasma density $N(r)$ is proportional
to $1/r^{2+A}$ where
$r$ is the distance from the center of Jupiter. The guess is that
$A\geq 0$, and we will assume that $A=0$ for the worst possible case. Hence,
radial dependence of the electron plasma density is taken as
$N(r)=N_0(R_J/r)^{2+A}$, where $R_J=7.1\times 10^7$ m is the mean radius of
Jupiter.

The plasma produces a delay $\Delta T$ in the time of propagation of radio
signal which is proportional to
the column plasma density in the line of sight given by integral
(\cite{yak})
\begin{eqnarray}
\label{p}
N_l&=&\int^{r_0}_d {N(r)\,dr\over r^{A+1}\sqrt{r^2-d^2}}+\int^{r_1}_d
{N(r)\,dr\over r^{A+1}\sqrt{r^2-d^2}}\;,
\end{eqnarray}
where $r_0$ and $r_1$ are radial distances of quasar and radio antenna from
Jupiter respectively, and $d=|{\bm \xi}|$ is the impact parameter of the
light ray from the quasar to Jupiter \footnote{This impact parameter $d\simeq
14R_J$ on September 8, 2002 and corresponds to the angle $\theta=3.7'$
between the quasar and Jupiter in the plane of the sky.}. In the experiment
under discussion the impact parameter is much less than both $r_0$ and $r_1$.
Hence,
\begin{eqnarray}
\label{pp}
N_l(\mbox{m}^{-2})&=&N_0R_J\left({R_J\over
d}\right)^{A+1}{\sqrt{\pi}\Gamma\left({A+1\over 2}\right)\over
\Gamma\left(1+{A\over 2}\right) }\;,
\end{eqnarray}
where $\Gamma(z)$ is the Euler gamma-function.
The plasma time delay
\begin{eqnarray}
\label{ppp}
\Delta T\mbox{(s)}&=&40.4\,c^{-1}\nu^{-2} N_l \;,
\end{eqnarray}
where $c$ is the speed of light in vacuum measured in m/sec, $\nu$ is the
frequency of electromagnetic signal measured in Hz.

It is worthwhile noting that, in fact, the VLBI array measures difference in
path length between the radio telescopes. Hence, one has to differentiate
$N_l$ in expression (\ref{ppp}) with respect to the impact parameter $d$ and
project the result on the plane of the sky. This gives a magnetospheric VLBI
time delay of
\begin{eqnarray}
\Delta_{JM}\mbox{(ps)}&=&6.3\times 10^{-7}(A+1){N_l\over
d}\left({\nu_0\over \nu}\right)^2{{\bm n}\cdot{\bm b}\over c}\;,
\end{eqnarray}
normalized to the frequency $\nu_0=8.0$ GHz. Substituting $d=13R_J$ and
taking the baseline $b=6000$ km we find
\begin{eqnarray}
\label{qqq}
\Delta_{JM}&=&2.34\left({\nu_0/\nu}\right)^2\; \mbox{ps}\qquad\quad(A=0)\;,
\\
\Delta_{JM}&=&0.13\left({\nu_0/\nu}\right)^2\; \mbox{ps}\qquad\quad(A=1)\;,
\\
\Delta_{JM}&=&0.03\left({\nu_0/\nu}\right)^2\; \mbox{ps}\qquad\quad(A=2)\;.
\end{eqnarray}
This represents the pure bending from Jupiter's magnetosphere which should be
compared
with the propagation of gravity time delay $\Delta_{JPG}=5.1$ ps at closest
approach.

If we observe at the dual frequencies at 2.3 GHz and 8.4 GHz in the normal
geodetic mode, we certainly can determine the ionospheric (both from
Jupiter and from the Earth) effects.  However, our sensitivity at
8.4 GHz will be decreased because of only one polarization and only
half the total bandwidth. The noise in the ionosphere/magnetosphere
bending may also be a limit. As a rough order of magnitude, the position
error per day is 10 $\mu$as at 8.4 GHz and 30 $\mu$as at 2.3 GHz.
Whatever bending we obtain at 2.3 GHz, about 10\% is
removed from the 8.4 GHz bending.  Thus, the ionosphere/magnetosphere
correction will have an error of 3 to 4 $\mu$as.

Only in the worst case scenario will the effects of
Jupiter's magnetosphere be significant at 8 GHz observing. Perhaps we should
observe at two widely spaced frequencies in the 8
GHz band, say at 8.0 GHz and 8.5 GHz, then, independently reduce the data at
the two frequencies.
The gravitational
bending delay (\ref{a6}) and gravitational retardation of gravity delay
(\ref{a7}) are both
independent of frequency.  Any plasma delay should scale
inversely with the frequency-squared and can be determined by looking at the
difference measurements. However, this small frequency difference will
produce very large errors in the estimate of the jovian bending and also be
strongly affected by Earth ionospheric contamination.

If we observe at 15 GHz instead, the magnetospheric and ionospheric delays
are both a factor
of four smaller and almost certainly negligible.  However, the system
sensitivity is less and one of the
calibrators may be too weak to reliable detect (J0839+1802).
We are still unsure about the most optimum method to deal with the possible
jovian magnetosphere component.

Time variability of the Jupiter magnetosphere could cause problems.  For
example, if there were very large, chaotic changes in the jovian
magnetosphere, then
we could lose coherence over a minute of time.
However, this fluctuation model is
very pessimistic and unlikely, and would probably average out to the
steady state model.

\section{Summary}

We believe that the differential VLBI experiment in September 2002 can
measure the retardation effect in propagation of gravity and determine
the speed $c_g$ of its propagation with 10\% to 20\% accuracy.  If the
experiment is successful it will provide a new independent test of
general relativity in the solar system.

\begin{acknowledgements}
This project has been partially supported by the University of
Missouri-Columbia Research Council grant URC-01-083. We thank B. Mashhoon and C.R. Gwinn for discussions and
valuable comments and A. Corman for help in preparation of the manuscript.
\end{acknowledgements}

\end{document}